\documentclass[12pt]{iopart}

\usepackage{graphicx}

\newcommand{\ep}{\epsilon}

\def\tr{\mbox{tr}}

\renewcommand{\>}{\rangle}
\newcommand{\<}{\langle}

\usepackage{color}

\begin{document}
\title{Fermi Edge Resonances in Non-equilibrium  States of Fermi Gases}

\author{E. Bettelheim}
\address{Racah Institute of Physics, Hebrew University, Jerusalem, Israel}
\author{ Y.Kaplan}
\address{Racah Institute of Physics, Hebrew University, Jerusalem, Israel}
\author{P. Wiegmann}
\address{The James Franck  Institute, University of Chicago}
\begin{abstract}
We formulate the problem of the Fermi Edge Singularity in non-equilibrium states of a Fermi gas   as a matrix Riemann-Hilbert problem with an integrable kernel.  This formulation is the most suitable for studying the singular behavior at each edge of non-equilibrium Fermi states by means of the method of steepest descent, and also reveals the integrable structure of the problem.  We supplement this result by  extending the familiar approach to the problem of the Fermi Edge Singularity via the bosonic representation of the electronic operators to non-equilibrium settings. It provides a compact way to extract the leading asymptotes.
\end{abstract}
\maketitle
\section{Introduction}
The FES  (Fermi edge singularity) \cite{NozieresDedominicis,Mahan,SchotteSchotte,Ohtaka:Tanabe}  is observed
in absorption of X-rays in metals as a power law peak at the Fermi Edge of a degenerate Fermi gas. There a sudden removal of a localized  electron from a hard core atomic shell creates a potential which disturbs the electronic gas, thus producing a power low spectrum of electronic soft modes.  In recent years the FES has also been demonstrated in  tunneling experiments \cite{Geim:FermiEdge, Cobden:Muzykantskii:Fermi:Edge, Hapke:Wurst, Larkin:Fermi:Edge:Mangetic:Induced}. There  a single electron can change the capacity of a contact  producing a similar disturbance to the electronic gas as a localized hole.  As a result, one observes a power law in tunneling current vs. the bias voltage:  $I(V)\sim V^{-2a+ka^2}$ \cite{Matveev:Larkin}, where $\delta=\pi a$ is the scattering phase of the ensuing potential and $k$ is the number of scattering channels.  In the case of an attractive potential $(a>0)$ the current  peaks at the Fermi edge.

One of the reasons of interest in the FES, and our own motivation in studying it, is that the origin of the FES may be found to be ascribable solely to Fermi statistics. It has been  studied over at least five decades, has been well understood,   and is considered  as one of the fundamental quantum phenomena in electronic physics. Early theoretical papers \cite{NozieresDedominicis,Mahan,SchotteSchotte} on the FES and the related phenomenon of Orthogonality Catastrophe \cite{Anderson:Catastrophe} were proved to be influential well  beyond FES. They are at the foundation of the modern physics of electronic systems in low dimensions.

In a degenerate Fermi gas anything but the leading power asymptotes is rarely of any interest.
 The reason for that is that everything else, except the leading power, depends on details on band structure, tunneling contacts, etc., and lacks of  universal character.

A different situation occurs in a non-equilibrium Fermi gas. There the energy scale of non-equilibrium features can be  much smaller than the Fermi scale and can be seen in the spectrum of absorption or tunneling. In this case, additional features become universal.

\begin{figure}[h!!!]
\begin{center}
\includegraphics[width=8cm]{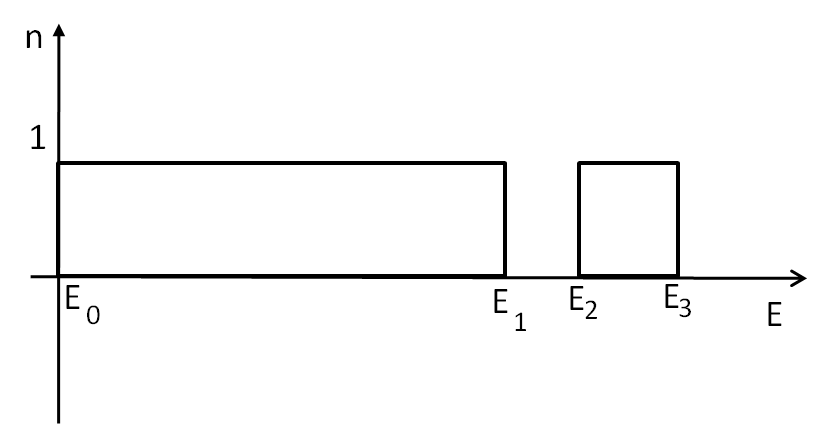}
\caption{Occupation of electronic states in structured Fermi sea. }\label{Fig1RH}
\end{center}
\end{figure}

As a prototype of a non-equilibrium state we consider a state where the Fermi distribution consists of steps at $E_{1}<E_{2}<\dots  E_{2n+1}$ such that no states are occupied in the interval between energies $E_{2i-1},E_{2i}$, where $i=1,\dots,n$. We denote by $E_0$ the bottom of the filled Fermi sea, which is assumed to be far away Fig.~\ref{Fig1RH}. Assuming that the scattering phase does not change within a wide range of the conducting band, say, between $E_0=0$ to $E_{2n}=\Lambda$
the spectrum will be a transcendental  universal function of $V/E_{ij}$, where $E_{ij}=E_i-E_j$.

These structured non-equilibrium states described above inevitably appear in the evolution of an arbitrary semiclassical Fermi state \cite{Bettelheim:Kaplan:Wiegmann:Bosons}. They were also realized in some  nanoscale devices (see e.g. \cite{wires}).
 \begin{figure}[b!!!]
\begin{center}
\includegraphics[width=10cm]{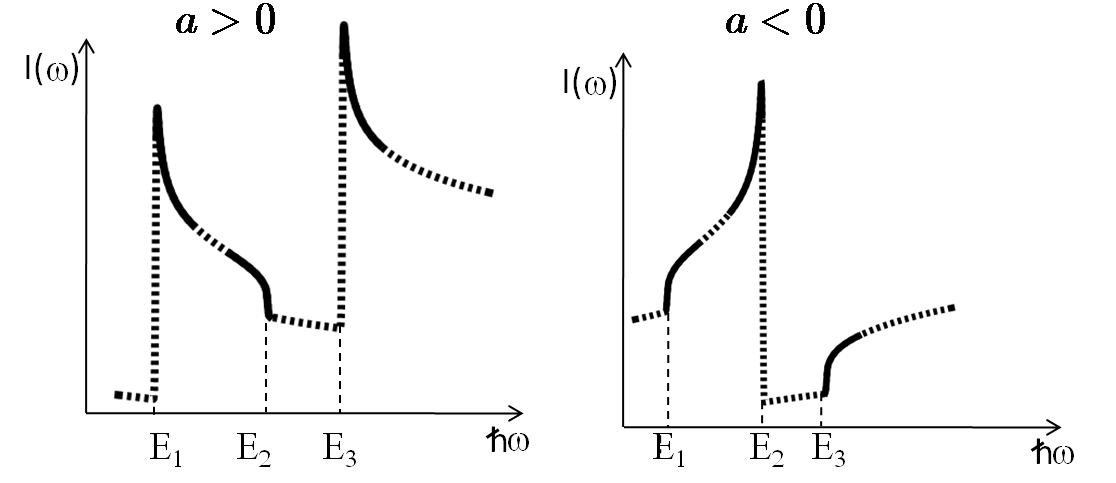}
\caption{A schematic plot of tunneling current for  small $a<0$ (left panel) and $a>0$ (right panel), solid lines show computed power law asymptotes.  Dashed lines interpolate between resonances.    \label{2}}
\end{center}
\end{figure}

In a non-equilibrium state  the absorption spectrum (or tunneling current)  is a transcendental function and elementary methods can determine its asymptotic expansion in various regimes.  In this paper we develop a framework aiming to characterize the universal part of the  spectrum. We derive a $2\times2$ matrix  Riemann-Hilbert problem derived from  an inversion problem of an integrable kernel. A similar matrix RH-problems appeared in  studies of various  fermionic correlation functions and correlation functions of eigenvalues of Random Matrices \cite{Deift:Its:Zhou:Riemann}.
One of a proven advantage of this formulation is that RH-problem is the most suitable for application of the steepest descent method. The latter prompts the leading asymptotes at Fermi edges.  Besides the fact that the RH-problem connects FES to a number of physically unrelated but mathematically equivalent  (often  well studied) problems, we think that these links
and analytical apparatus  they bring to physics of the FES  are important for a deeper understanding quantum non-equilibrium phenomena.  A somewhat alternative, but essentially equivalent approach is to establish a set of non-linear integrable equations \cite{Bettelheim:Abanov:Wiegmann:QHydrodynamics:JPHYSA}. We do not discuss this approach here.

As far as the main asymptote is concerned,  we will show that in the setting described above the tunneling current at voltage close to a Fermi edge $ \varepsilon_i\left(eV-E_{i} \right) \ll |E_{i,i\pm 1}|$ reads:
\begin{eqnarray}
\label{result1}
I(V)\propto A_i \left| eV-E_{i}\right |^{(2n-1)a^2-2\varepsilon_ia},\\
A_i=\mbox{const}\,\prod_{l< m}^{2n} E_{lm}^{\,2 \varepsilon_n\varepsilon_m a^2}\prod_{0\leq j\neq i}^{2n}E_{ij}^{-2\varepsilon_j a},
\end{eqnarray}
where  $\varepsilon_i=\pm 1$, if $E_i$ is an lower/upper edge of an occupied band, respectively.

Some noticeable features of this result are: (i) the exponent depends on the total number of bands, but stays the same for all upper (lower) edges of occupied intervals $\frac{d I}{IdV}= (2na^2-(a-\varepsilon_i)^2)|V-E_{i}/e|^{-1}$; (ii) if the potential is attractive $a>0$ (which is the common case) the current features a peak with a power low decay at upper  edges  towards increasing energy; and in contrast  the current is suppressed at lower edges; if the potential is repulsive, $a<0$, and sufficiently small, a peak appears at the edge $E_{2k}$ with a power law to the left to the edge  Fig.\ref{2};
(iii) the logarithm of the factor $A_i$ can be treated as the electrostatic energy of particles with alternating charges $\pm a$ positioned at the upper/lower edges with an insertion of a unit $\epsilon_i$ charge at the edge $E_i$.

Fermi Edge singularities with a structured Fermi distribution have been studied before. In 1984 Combescot and Tanguy \cite{Combescot:Tanguy:I, Combescot:Tanguy:II, Combescot:Tanguy:III} considered a situation where the interval $(E_1,E_2)$ of the band $(0,\Lambda)$ is occupied, while the interval $(0,E_1)$ starting from the bottom of the band is unoccupied. Later 2004 the Abanin and Levitov \cite{Levitov:Abanin}  considered FES with a two steps Fermi distribution. A more general situation has been considered in the recent papers by Gutman et al \cite{Gutman:Gefen:Mirlin:Toeplitzd}, and by the authors \cite{Bettelheim:Kaplan:Wiegmann:Bosons}, where Eq. (\ref{result1}) have been obtained. The approach employed in these papers (with the exception of Ref.\cite{Bettelheim:Kaplan:Wiegmann:Bosons}) is based on the expression of the tunneling current as a determinant of  a Fredholm operator. A basis of particle-hole excitations of the structured Fermi sea has  been used to write the Fredholm operator in early papers  \cite{Combescot:Tanguy:I, Combescot:Tanguy:II, Combescot:Tanguy:III}. This approach  has been developed by Othaka and Tanebe \cite{Ohtaka:Tanabe} in 1984. Contrary one-particle basis in an empty vacuum has been  used in later  articles \cite{Levitov:Abanin, Gutman:Gefen:Mirlin:Toeplitzd}. Naturally the basis of of particle-hole excitations of the Fermi sea, employed in  earlier papers  \cite{Combescot:Tanguy:I, Combescot:Tanguy:II, Combescot:Tanguy:III,Ohtaka:Tanabe}, which already captures the many-body physics of the Fermi sea is a step forward in obtaining the desired results.  We employ this approach here. In addition, for illustrative purposes (Appendix \ref{BosonicAppendix}), we also present a simple a compact method to capture a leading singularity  developed in Ref.\cite{Bettelheim:Kaplan:Wiegmann:Bosons}. That method is  based on the bosonic representation of electronic operators in a Fermi state with multiple edges.

\section{Tunneling current.} The tunneling  setting of FES is as  follows:
a Fermi gas is in contact  with  a localized resonant level (a quantum dot). It is initially uncharged and provides no scattering to electrons.
When an electron tunnels to the dot, it  suddenly charges the dot, switching-on a  small potential $H \to H'=H+U$
  localized at the dot  \cite{Matveev:Larkin}.
We assume no further  interaction, no dissipation, and we ignore spin and channels.

The tunneling current is given by the golden rule \cite{NozieresDedominicis,Matveev:Larkin}. In units of the tunneling amplitude,  $I(\omega)|_{\hbar\omega=eV+E_1}$  reads
\begin{eqnarray}
 I(\omega)\propto \mbox{Re}  \int_0^\infty e^{i\omega(t_1-t_2)}   G(t_1,t_2) d\tau,\\
 \label{red}
G(t_1,t_2) =  \<\Omega|e^{iHt_2}c e^{iH'(t_{1}-t_2)}
 c^\dag\; e^{-iHt_1}|\Omega\>
\end{eqnarray}
Here $c=\sum_\epsilon c_\epsilon$ is an electronic operator in the position of the dot, assumed to be the origin, while $c_\epsilon$ is an electronic mode with energy $\epsilon$ and $|\Omega\>$ is the structured Fermi state of interest. We also assume that the bias voltage is with respect to $E_1$.

Next we assume that the potential is regular within a wide energy range which exceeds the energy  range of features of the structured Fermi state and that the tunneling time is sufficiently small. Under this assumption the energy dependence of the scattering phase $\delta$ caused by potential $U$
can be dropped. This amounts to a  downwards shift of the energy levels by a constant amount  $a$ (in units of level spacing):  $\epsilon\to \epsilon-a$.

In Ref. \cite{SchotteSchotte} it has been shown that  the vertex operator $e^{a\varphi}$ implements a  shift of momenta: the perturbed Hamiltonian and perturbed states
are seen as similarity transformation of the unperturbed ones
$H' =  e^{-a \varphi } H e^{ a\varphi}$ and $|\Omega'\>=e^{a\varphi}|\Omega\>$.
Here an operator $\varphi$  is a chiral canonical Bose field related to the chiral part of electronic density
\begin{equation}\label{B}
\varphi(t)=\hbar\sum_{\epsilon\neq 0}  e^{\frac{i}{\hbar}\epsilon t}\rho_\epsilon/\epsilon, \quad \rho_\epsilon=\sum_{\varepsilon} c^\dag_{\varepsilon}c_{\varepsilon+\epsilon}.
\end{equation}

 Then Green's function reads
\begin{eqnarray} \label{G}
G(t_1,t_2) =  \<\Omega|c(t_2)
e^{ - a\varphi(t_2) }e^{  a\varphi(t_1)} c^\dag(t_1)|\Omega\>,
\end{eqnarray}
where $c(t)=\sum_\epsilon e^{\frac i\hbar \epsilon t}c_\epsilon$. This formula is standard.

\section{Fredholm Determinants} Following \cite{NozieresDedominicis}, Green's function can be understood as consisting of  three multiplicative factors $G(t_1,t_2)=|\< \Omega'| \Omega\>|^2 e^{C}\cdot L$ - an overall normalization:
\begin{equation}
|\< \Omega'| \Omega\>|^2, \quad \mbox{where}\quad  |\Omega'\> = e^{a \varphi(0)}|\Omega\>,
\end{equation}  closed loops
\begin{equation}
e^C=\frac{\<\Omega| e^{ - a\varphi(t_2) }e^{  a\varphi(t_1)} |\Omega\>}{|\< \Omega'| \Omega\>|^2},
\end{equation}
and open lines
\begin{equation}
L=\frac{\<\Omega|c(t_2)
e^{ - a\varphi(t_2) }e^{  a\varphi(t_1)} c^\dag(t_1)|\Omega\>}{\<\Omega|
e^{ - a\varphi(t_2) }e^{  a\varphi(t_1)} |\Omega\>}.
\end{equation}
The two latter objects can be cast in the form of Fredholm determinant  by means of the
Wick theorem. We remind the major formulas.

Consider coherent states of $Gl(\infty)$. These states are obtained by transforming the ground state  of the Fermi gas $|A\>=g(A)|0\>$ by an exponent of a bilinear form of Fermi operators $g(A)=e^{\sum_{\epsilon\eta}A_{\epsilon,\eta}c^\dag_\epsilon c_{\eta}}$, where $A_{\epsilon,\eta}$ is an arbitrary  $gl(\infty)$ matrix. Our structured Fermi state is a coherent state.

For arbitrary coherent states $A_1,A_2$ and arbitrary  $gl(\infty)$ matrices $B,C$ the following holds
\begin{equation}\label{det}
	\frac{\<A_1|g(B)g(C)|A_2\>}{\<A_1|g(B)|A_2\>\<A_1|g(C)|A_2\>}=\det ({\bf 1+ K}),
\end{equation}
where
\begin{equation}\label{K}
K_{\ep_1,\ep_2}= \sum_{\ep}M_{\ep_1,\ep}(A_1,B,A_3)M^\dag_{\ep,\ep_2}(A_3,C,A_2)
\end{equation}
and
 \begin{equation}\label{M}
M_{\ep,\eta}(A_1,B,A_3)=\< A_1|g(B) c^\dagger_{\ep} c_{\eta} | A_3 \>
\end{equation}
are matrix elements of operator $g(B)$ between the state $\< \Omega(A)|$ and a state $c^\dagger_{\ep} c_{\eta}|A_3\>$,  where a  particle-hole pair is added to an arbitrary chosen coherent  state $|A_3\>$. The result does not depend of the choice of $A_3$.

In order to obtain this formula one inserts a superposition of an arbitrary number of particle-hole excitations into a chosen coherent state $|A_3\>\<A_3|$, apply the Wick theorem to each term and sum them up.

The variation of (\ref{det}) $\delta \log\det ({\bf 1+ K})=\tr\left[ ({\bf 1+ K})^{-1} {\bf \delta K} \right]$ gives another known  formula
\begin{equation}\label{vardet}
	\frac{\<A_1|c^\dag_{\ep_1}g(B)g(C)c_{\ep_2}|A_2\>}{\<A|g(B)g(C)|A'\>}=\tr\left[ ({\bf 1+ K})^{-1} {\bf P} \right],
\end{equation}
where
\begin{equation}\label{P}
P_{\ep,\eta}(\ep_1,\ep_2) =\tilde M_{\ep_1\eta}(A_1,B,A_2)\tilde M^*_{\eta,\ep_2}(A_1,C,A_2),
\end{equation}
and $\tilde M_{\ep\eta}(A_1,B,A_2)=\< A_1| c_{\ep}g(B) c^\dagger_{\eta} | A_2 \>$.
Specification of these formulas: $\<A_1|=\<A_2|=\<\Omega|,\, g(B)=g^\dag(C)=e^{-a\varphi(t)}$ prompts a determinantal representation of the current
\begin{eqnarray}\label{D}
e^C=\det ({\bf 1+ K}),\; L=\tr\left[ ({\bf 1+ K})^{-1} {\bf P} \right]
\end{eqnarray}
with a  kernel
\begin{eqnarray}\label{KandP}
K({\ep_1,\ep_2})  =  \sum_{\eta}M_{\ep_1,\eta}(t_1)M^*_{\eta,\ep_2}(t_2),
\\
P_{\ep,\eta} =M_{\ep}(t_1)M^*_{\eta}(t_2),
\end{eqnarray}
where
\begin{eqnarray}
 M_{\ep,\eta}(t)=\frac{\< \Omega|e^{-a\varphi(t)} c^\dagger_{\ep} c_{\eta} | \Omega \>}{\<\Omega'|\Omega\>}, \label{MDefinition}\\
 M_{\ep}(t)=\frac{\< \Omega | c(t) e^{-a\varphi(t)}  c^\dag_{\ep} |\Omega \>}{\<\Omega'|\Omega\>}
\end{eqnarray}
 are matrix elements of the vertex operator between  states where a  particle-hole pair, or just one particle are added to the state $|\Omega\>$ \footnote{Formulas equivalent to  (\ref{KandP})  for the ground state can be found  in \cite{Ohtaka:Tanabe}.}. These are general formulas valid for any coherent state $|\Omega\>$.

The formulas is further specified since the structured Fermi state we are considering is  an eigenvalue of the Hamiltonian.
In this case
\begin{eqnarray}\label{M1}
M_{\ep\eta}(t)=e^{\frac{i}{\hbar} (\eta-\ep)t} \frac{\< \Omega'|\Omega;\ep,\eta \>}{\<\Omega'|\Omega\>},\\
M_{\ep}(t)=e^{-\frac{i}{\hbar} \ep t}  \frac{\< (\ep;\Omega)'|\Omega;\ep \>>}{\<\Omega'|\Omega\>} \label{M2}.
\end{eqnarray}
$M_{\ep\eta}$ is an   overlap between the state $\<\Omega'|$ which appears after the shake-up and a particle-hole excitation of the state $|\Omega\>$ before shake-up with energy $\ep\notin\Omega$ and $\eta\in\Omega$, where we denote $\Omega=\cup_{i=0}^{n-1}(E_{2i},E_{2i+1})$ as the set of  occupied single particle states in $|\Omega\>$, $(\Omega,\ep)$ and $(\Omega;\ep,\eta)$ are states where an extra particle or a particle-hole pair is added into $|\Omega\>$.  Similarly  $M_{\ep}$  is the overlap of states $|\Omega \>$ and the state $|\Omega'\>$ with an added  particle with energy $\ep$.
\section{Matrix elements}
The following formula helps evaluating the matrix elements (\ref{M1},\ref{M2}):
If $|\ep'\>$ is a single particle eigenstate of the perturbed Hamiltonian and $|\ep\>$ is a one-particle eigenstate of an unperturbed state then their overlap (in units of  level spacing)  is  $\<\ep'|\ep\>=\frac{\sin(\pi a)}{\pi(\ep-\ep')}$. An extension of this formula  to many particle states $\<\underline \ep|=\<\ep_1,\dots, \ep_N|$ and $|\underline \ep'\>=|\ep_1',\dots,\ep'_N\>$ gives  a Cauchy determinant
\begin{equation}\label{Cauchy}
\left(\frac{\pi}{\sin(\pi a)}\right)^{N} \<\underline \ep|\underline \ep'\>=
 \det\frac{1}{\ep_i-\ep'_j}=
\frac{\prod_{i>j}(\ep_i-\ep_j) (\ep'_i-\ep'_j)}{ \prod_{i,j} (\ep'_i-\ep_j)}.
\end{equation}
With the help of  this formula, the matrix elements in (\ref{KandP}) can  be computed in a manner  similar to Ref. \cite{Ohtaka:Tanabe}. While computing one must take into account that the  set of occupied single particle levels in  $\Omega$ and $\Omega'$ are shifted  by $a$ with respect to each other. Then the problem is reduced to an electrostatic problem of placing a dipole or a charge into a Coulomb plasma confined in the intervals $\cup_{i=1}^{n}(E_{2i}, E_{2i+1})$.  It gives the overlap between states $\<\Omega|$ and $|\Omega'\>$ generalizing  Orthogonality Catastrophe formula \cite{Anderson:Catastrophe}. Up to an $E_i$ -independent constant factor it reads
\begin{eqnarray}\label{normalization}\<\Omega|\Omega'\>\sim \Delta^{(n+1) a^2}\prod_{i>j}E_{ij}^{\epsilon_i\epsilon_j a^2}.
\end{eqnarray}
The results for matrix elements are
\begin{eqnarray}\label{MM}
M_{\ep\eta}=e^{\frac{i}{\hbar} (\eta-\ep)t}\frac{r(\ep)s(\eta)}{\ep-\eta},\;M_\ep=e^{\frac{i}{\hbar} \ep t}r(\ep),\\
r(\ep)=\prod_{i=1}^{2n} \left(\ep-E_{i}\right)^{\varepsilon_ia},\quad \ep\notin\Omega; \\
s(\eta)=\frac{\sin\pi a}{\pi}\prod_{i=1}^{2n} \left(E_{i}-\eta\right)^{-\varepsilon_ia},\quad \eta\in\Omega
\nonumber
\end{eqnarray}
where $\ep\notin\Omega,\;\eta\in\Omega$ are energies of particles and holes. A short sketch  of these calculations is found in Appendix \ref{CauchyAppendix}.

Summing up, the kernel reads
\begin{eqnarray}
K(\ep_1,\ep_2)=e^{\frac{i}{\hbar}(\ep_2 t_2-\ep_1 t_1)}r(\ep_1)r(\ep_2)
\frac{Q(\ep_2)-Q(\ep_1)}{\ep_1-\ep_2},\label{KK}\\
Q(\ep,\tau) =  \int_{{\eta\in}\Omega} e^{\frac{i}{\hbar}\eta \tau}\frac{s^2(\eta) d\eta}{\ep-\eta},\quad \tau=t_1-t_2.\label{Q}
\end{eqnarray}

\section{Integrable kernel} The next step is to invert the Fredholm kernel $\bf K$. It can be done in a straightforward  manner similar to \cite{Ohtaka:Tanabe,Combescot:Tanguy:I, Combescot:Tanguy:II, Combescot:Tanguy:III} employing the Wiener-Hopf method at every edge. However, calculations become more structured if we use the {\it integrable}  property of  the kernel. Integrability is  general property  of free fermion correlators (see \cite{Deift:Zhou:Steepest:Descent,Sato:Miwa:Jimbo:Impenetrable,Its:Korepin:DiffEqForCorr,Deift:Its:Zhou:Riemann, Bettelheim:Abanov:Wiegmann:QHydrodynamics:JPHYSA}.

A kernel is called {\it integrable} if it has the form
\begin{eqnarray}\nonumber
K(\ep_1,\ep_2)= \frac{\sum_{\alpha=1}^l f_\alpha(\ep_1) g_\alpha (\ep_2)}{\ep_1-\ep_2}, \quad \sum_{\alpha=1}^l f_\alpha(\ep)g_\alpha(\ep) =0.
\end{eqnarray}
In the case of a structured Fermi sea $l=2$, and,  as follows from  (\ref{KandP},\ref{MM}):
\begin{eqnarray}
g_1(\ep)=Q(\ep,\tau)g_2(\ep),\quad f_2 (\ep)=-Q(\ep,\tau)f_1(\ep),\label{fg}\\
g_2(\ep)=e^{\frac{i}{\hbar}\ep t_2}r(\ep),\quad f_1(\ep)=e^{-\frac{i}{\hbar} \ep t_1}r(\ep) ,\nonumber
\end{eqnarray}
Let  $\vec F=({\bf 1+K})^{-1}{ \vec f}$ (we  denote $\vec f=(f_1,f_2)$)
be a solution of the singular integral equation
\begin{eqnarray}\label{F}
\vec F(\ep_1)+\int_{\ep_2\notin\Omega}K(\ep_1,\ep_2)\vec F(\ep_2)d\ep_2=\vec f(\ep_1),\quad \ep_1\notin\Omega.
\end{eqnarray}
The time derivative of  closed loops contribution  and a contribution of  open lines (\ref{D})  are expressed through the solutions $\vec F=(F_1,F_2)$
\begin{eqnarray}\label{CL1}
\frac{dC}{d \tau}=\tr\left(({\bf 1+K})^{-1}\frac{d{\bf K}}{d \tau}\right)=\frac{i}{\hbar}\int_{\ep\notin\Omega}(g_1 F_1 - g_2 F_2) d\ep\\
\label{OL}
 L=\int_{\ep\notin\Omega} g_2 F_1 d\ep,
\end{eqnarray}
\section{Matrix Riemann-Hilbert problem}    The Fredholm equation (\ref{F})  is sufficient to obtain the singular behavior at  Fermi edges. However,
it is instructive to cast the FES problem as a matrix RH problem along the lines described in \cite{Deift:Its:Zhou:Riemann}.  In that form, the FES problem falls in the general scheme of integrable problems.  In addition, the RH-problem  is the   most suitable  for  analysis near edges  \cite{Deift:Zhou:Steepest:Descent}.

The central object of the RH problem is a matrix-valued functions $m(\ep)$  analytic in a complex $\ep$-plane cut along the unoccupied intervals $\cup (E_{2i-1},E_{2i})$ Fig.~\ref{Contour}, defined such that at infinity $m$ approaches the unit matrix, and that its  boundary value on the cuts  $m_\pm=m (\ep\pm i0)$ connects vector the $\vec F$ to the vector $\vec f$ as
\begin{eqnarray}
\vec F(\ep)=m_+(\ep)\vec f(\ep),\quad \ep\in\Omega.
\end{eqnarray}
In Ref. \cite{Deift:Its:Zhou:Riemann} it has been shown that the matrix is a solution of the RH-problem:
\begin{eqnarray}\label{m}
m_+ v=m_- ,\quad v_{\alpha\beta}= \delta_{\alpha\beta}-2\pi if_\alpha g_\beta.
\end{eqnarray}

In the case of FES
\begin{eqnarray}\label{v}
v(\ep) =\mathbf{1}+2\pi ie^{-\frac{i}{\hbar} \ep\tau}r^2(\ep)\left(\begin{array}{cc}Q & 1 \\-Q^2 & -Q\end{array}\right).
\end{eqnarray}
Eqs (\ref{CL1}-\ref{v}) constitute the matrix RH-problem for FES. As typical for other integrable RH-problems, a similarity transformation can be found to reduce the jump matrix $v(\ep)$ to a constant matrix, such that analytic behavior  in the energy dependence of the kernel will be translated to into the analytic nature of singularities of the solution at infinity. We do not do this here.

Being specified for a one-edge problem  ($n=0$), in units of upper and lower cut-offs, read
\begin{eqnarray}
r^{(0)}(\ep)= \ep^{\varepsilon_k a},\quad \varepsilon_k\ep>0;\\
Q^{(0)}(\ep,\tau)=\left(\frac{\sin\pi a}{\pi}\right)^2 \int_{\varepsilon_k\eta<0} e^{\frac{i}{\hbar}\eta \tau}\frac{ (-\eta)^{-2\varepsilon_ka} d\eta}{\ep-\eta},
\end{eqnarray}
where we count  energy from the edge. In this case the RH-problem is solved by elementary means. In fact technically it easier to proceed directly through the integral equation following \cite{Ohtaka:Tanabe}.
\section{Method of steepest descent and the leading singularity}
The asymptotic behavior at the edges can be found by the steepest-descent method described in \cite{Deift:Zhou:Steepest:Descent}.
\begin{figure}[h!!!]
\begin{center}
\includegraphics[width=4cm]{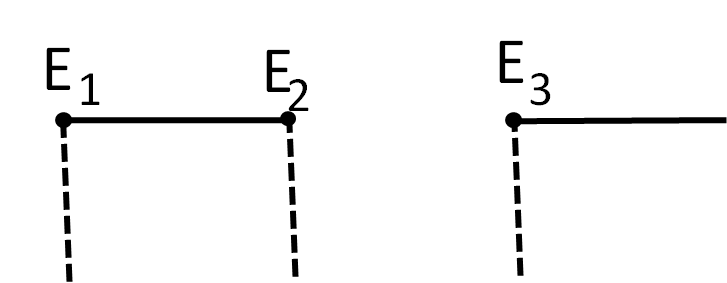}
\caption{Unoccupied electronic states are between edges  $E_{2k-1}$ and $E_{2k}$, $k=1,\dots n$.
The matrix $m$  of the RH-problem  (\ref{m})  jumps on segments of real axis corresponding to unoccupied states (solid  line). The steepest descent contour goes down vertically (dashed line) in the lower half plane.}\label{Contour}
\end{center}
\end{figure}

The steepest-descent contour starts from Fermi edges and extends to $i\infty$ in the lower half-plane as is in Fig.~\ref{Contour}. Along this contour the rapidly falling exponential factor $e^{-\frac{i}{\hbar}\ep \tau}$ in (\ref{v}) suppresses  the jump of the matrix $m$ except at small segments near the Fermi edges, where $r(\ep)$ and $Q(\ep)$ are singular. At energy close to an edge, say, $E_i$ we estimate $r(E_i+\ep)\approx r^{(0)}(\ep) \prod_{j\neq i}^{2n} E_{ij}^{\varepsilon_ia}$ and $Q(E_i+\ep,\tau) \approx e^{\frac{i}{\hbar}E_i \tau}\prod_{j\neq i }^{2n} E_{ji}^{-2\varepsilon_ia}Q^{(0)}(\ep),$ and the problem reduces by a similarity transformation to the one-edge  problem
\begin{eqnarray}
\vec{f}(E_i+ \ep) ={\cal Z}_i \vec{f}^{(0)}(\ep);\quad \vec{g}(E_i+ \ep) ={\cal Z}_i^{-1} \vec{g}^{(0)}(\ep);   \nonumber\\
m(E_i+\ep) = {\cal Z}_i m^{(0)}(\ep) {\cal Z}_i^{-1}; \nonumber\\
{\cal Z}_i=e^{-\frac{i}{2\hbar} E_i (\mathbf{1}\cdot t - \sigma_3 \tau) } \prod_{j\neq i} |E_{ij}|^{\sigma_3\varepsilon_j a},
\end{eqnarray}
where script $(0)$ indicates entities of the one-edge problem, $t=t_1+t_2$, and $\mathbf{1}$ is the $2\times2$ unit matrix. 

The reduction of the multi-edge problem to the one edge problem using the RH steepest descent method is what allows us to solve the problem. Indeed, the contribution of each edge is well known. In units of spacing and up to a constant factor  they are
\begin{eqnarray}\label{38}
C^{(0)}_i\sim \tau^{-a^2}, \quad L^{(0)}_i\sim \varepsilon_i \tau^{2a-1}.
\end{eqnarray}
We note that Ohtaka and Tanebe \cite{Ohtaka:Tanabe} showed how (\ref{38})  originally obtained in Refs.  \cite{NozieresDedominicis,Mahan} by different means) follow from the integral equation (\ref{F}).

The leading asymptote for the structured Fermi sea can be obtained by a combination of the one edge problem and a similarity transformation.
The similarity transformation does not affect the contribution of closed loops. Therefore each edge contributes  equally.
Summing them up, we obtain $e^C \approx e^{\sum_i C_i^{(0)}}$.

The similarity transformation and (\ref{OL}) give the contribution of open lines as a sum of one edge open lines each weighted by its own amplitude
\begin{eqnarray}\label{L}
L=\sum_{i=1}^{2n+1}e^{-\frac i\hbar E_{i}\tau}\prod_{j\neq i}E_{ij}^{-2\varepsilon_ia}L_i^{(0)},
 \end{eqnarray}Combining the normalization the closed loop and the open line we obtain:
\begin{eqnarray}\label{GG}
G(\tau) \sim \tau^{-(2n+1)a^2}  \left(\prod_{i<j} E_{i,j}^{2 \varepsilon_i \varepsilon_j a^2}\right)   e\sum_m \varepsilon_m\tau^{2a-1} e^{-\frac{i}{\hbar} E_m \tau} \prod_{n\neq m} E_{m,n}^{-2 \varepsilon_m a},
\end{eqnarray}
where  a multiplicative constant depending only on cutoffs has been omitted. Fourier transforming, this formula prompts  the main result (\ref{result1}).
\section{Acknowledgment}
The authors acknowledge discussions with A. Mirlin, D. Gutman, S.-Y. Lee and especially A. Abanov.  P. W. was supported by NSF DMR-0906427, MRSEC under DMR-0820054. E. B. was supported by grant 206/07 from the ISF.

\appendix
\section{Compnutation of (\ref{normalization}-\ref{MM}) \label{CauchyAppendix}}

We sketch the computations  of matrix element (\ref{normalization}).  They are not much different from similar calculations for a single edge problem \cite{Ohtaka:Tanabe} . We start from a  general formula (\ref{Cauchy}) describing an overlap of arbitrary  electronic states consisting of a finite number $N$ of occupied levels. It can be seen as  the exponent of the electrostatic interaction energy  of log-interacting particles with  charges of size $+a$ at  $\epsilon_i \in \Omega$'s and $-a$ at $\epsilon'_i\in \Omega-a$. Thus we have a regular alternating pattern of charges, the negative charges at an offset of $-a$ with respect to the positive ones, the patten starts at $E_0$ and persists up to $E_{2n+1}$, with gaps at $[E_{2i-1},E_{2i}]$, $i=1,\dots,n$. The electrostatic energy is:
\begin{eqnarray}
 \sum_{i\neq j, \epsilon_i,\epsilon_j\in\Omega} \log \left(\frac{\epsilon_i-\epsilon_j}{\epsilon_i-\epsilon_j+  a}\right)-N\log a.
\end{eqnarray}
This sum can be evaluated in terms of Barnes functions, but we are interested only in the limit of large number of levels. There  we may think in terms of a density of dipoles with a polarization vector  $a$. The energy of this system is equivalent to energy of charges  $ \varepsilon_i a$ placed at edge $E_i$. The latter (up to $a$-dependent constant) is $ \sum_{i <  j} \varepsilon_i \varepsilon_j a^2 \log|E_i-E_j|$. The  exponent of the energy yields   (\ref{normalization}).

The Configuration of charges corresponding to  $M_{\epsilon,\eta}$ is a set of dipoles discussed above minus  a charge $+1$ at $\epsilon\in\Omega$  plus a charge $-1$ at  $\eta\notin\Omega$. The energy acquires the interaction energy  between charges at $\eta$ and $\ep$  and  charges $\pm a$ sitting at edges $E_i$. This addition is $ \sum_i \varepsilon_i a (\log|\epsilon- E_i| -\log |\eta- E_i|)$ plus an $a$ dependent constant. The origin of the constant is the energy of dipoles in the vicinity of a hole at $\epsilon$. It is
$\sum_{\epsilon_i \in \Omega}\log \frac{\epsilon_i-\eta-a}{\epsilon_i-\eta}$. Assuming that $\eta$ is far from edges the contribution goes from levels close to $\epsilon$. It gives $\log\left(\frac{\sin(\pi a)}{\pi}\right)$. All together it yields (\ref{MM}). Computation of $M_\ep$ is similar. In that case only interactions between the extra charge at $\ep\notin\Omega$ and residual charges at edges contribute. It does not incur a constant factor.


\section{Computation through bosonic representation \label{BosonicAppendix}}
The leading singularity can be  understood using a bosonic formalism. First we separate fast oscillatory modes at each edge
 $$c(t,x_0)=\sum_i e^{\frac{i}{\hbar}E_{i}t} \psi_i(t).$$
 Then we represent slow  modes  through components of the Bose field  $\partial_x \varphi_i = i \psi^\dag_i(t)\psi_i(t)$ as
\begin{eqnarray}\label{FermionOperator}
\psi_i\propto (\varepsilon_i \prod_{j\neq i}E_{ij}^{\varepsilon_j})^{1/2}e^{-\varepsilon _i \varphi_i}.
\end{eqnarray}
The Bose field (\ref{B}) is a sum of its components
$\varphi=\sum_k\varphi_i$.  Components of the Bose field represent particle-holes excitations close at each edge. At $\hbar/\tau\gg E_{ij}$  they can be treated as independent   canonical Bose fields.
Their variances $D_i(t_1,t_2)=-\frac 12\<\Omega|\left(\varphi_i(t_2)-\varphi_i(t_1)\right)^2|\Omega\>$   are not difficult  to compute. As follows form (\ref{B}), $D_{2k-1}$   are sums of  $(\cos\epsilon\tau-1)/\epsilon$ over all possible energy   of a particle-hole excitations provided that a particle is placed in the "gap" $(E_{2k-1},E_{2k})$ . Similarly $D_{2k}$ is  the sum over energies  of a hole-particle  excitations provided that a hole is placed to the band $(E_{2k},E_{2k+1})$.   Computing these integrals at $\tau \gg\hbar/|E_{ij}|$ one obtains
\begin{eqnarray}\label{CC}
D_i(\tau)=-\log\tau+\varepsilon_i\sum_{j\neq i}{\varepsilon_{j}}\log |E_{ij}|
\end{eqnarray}
The time independent term in (\ref{CC}) explains the  prefactor in (\ref{FermionOperator}):  the correlator $\<\psi^\dag_i(t_1)\psi_i(t_2)\>\propto
\frac{\varepsilon_i}{\tau}$ then also has to be obtained for the Bose field. This yields $( \prod_{j\neq i}E_{ij}^{-\varepsilon_i \varepsilon_j})e^{D_i}$.

In the Bose representation, Green's function (\ref{G}) is a sum of edge components
\begin{eqnarray}\label{G0}
 G(t_1,t_2)=\sum_i (\varepsilon_i \prod_{j\neq i}E_{ij}^{-\varepsilon_i \varepsilon_j})e^{\frac i\hbar E_{i}\tau}G_i(t_1,t_2),\\
 \nonumber
 G_i=\<e^{(\varepsilon_i-a)(\varphi_i(t_2)-\varphi_i(t_1))}\>
\prod_{j\neq i}\<e^{-a(\varphi_j(t_2)-\varphi_j(t_1))}\>
 \end{eqnarray}
Computing this, we obtain Green's function (\ref{GG}).

\section*{References}
\bibliographystyle{unsrt}
\bibliography{mybib}

\begin{thebibliography}{10}

\bibitem{NozieresDedominicis}
P.~{Nozi{\`e}res} and C.~T. {de Dominicis}.
\newblock {Singularities in the X-Ray Absorption and Emission of Metals. III.
  One-Body Theory Exact Solution}.
\newblock {\em Physical Review}, 178:1097--1107, 1969.

\bibitem{Mahan}
G.~D. {Mahan}.
\newblock {Excitons in Metals: Infinite Hole Mass}.
\newblock {\em Physical Review}, 163:612--617, 1967.

\bibitem{SchotteSchotte}
K.~D. {Schotte} and U.~{Schotte}.
\newblock {Tomonaga's Model and the Threshold Singularity of X-Ray Spectra of
  Metals}.
\newblock {\em Physical Review}, 182:479--482, 1969.

\bibitem{Ohtaka:Tanabe}
K.~{Ohtaka} and Y.~{Tanabe}.
\newblock {Theory of the soft-x-ray edge problem in simple metals: historical
  survey and recent developments}.
\newblock {\em Reviews of Modern Physics}, 62:929--992, October 1990.

\bibitem{Geim:FermiEdge}
A.~K. {Geim}, P.~C. {Main}, N.~{La Scala}, Jr., L.~{Eaves}, T.~J. {Foster},
  P.~H. {Beton}, J.~W. {Sakai}, F.~W. {Sheard}, M.~{Henini}, G.~{Hill}, and
  M.~A. {Pate}.
\newblock {Fermi-edge singularity in resonant tunneling}.
\newblock {\em Physical Review Letters}, 72:2061--2064, March 1994.

\bibitem{Cobden:Muzykantskii:Fermi:Edge}
D.~H. {Cobden} and B.~A. {Muzykantskii}.
\newblock {Finite-Temperature Fermi-Edge Singularity in Tunneling Studied Using
  Random Telegraph Signals}.
\newblock {\em Physical Review Letters}, 75:4274--4277, December 1995.

\bibitem{Hapke:Wurst}
I.~{Hapke-Wurst}, U.~{Zeitler}, H.~{Frahm}, A.~G.~M. {Jansen}, R.~J. {Haug},
  and K.~{Pierz}.
\newblock {Magnetic-field-induced singularities in spin-dependent tunneling
  through InAs quantum dots}.
\newblock {\em Phys. Rev. B}, 62:12621--12624, November 2000.

\bibitem{Larkin:Fermi:Edge:Mangetic:Induced}
Y.~N. {Khanin}, E.~E. {Vdovin}, L.~{Eaves}, I.~A. {Larkin}, A.~{Patane}, O.~N.
  {Makarovski{\u i}}, and M.~{Henini}.
\newblock {Magnetic-field-induced Fermi-edge singularity in the tunneling
  current through an InAs self-assembled quantum dot}.
\newblock {\em Soviet Journal of Experimental and Theoretical Physics},
  105:152--154, July 2007.

\bibitem{Matveev:Larkin}
K.~A. Matveev and A.~I. Larkin.
\newblock Interaction-induced threshold singularities in tunneling via
  localized levels.
\newblock {\em Phys. Rev. B}, 46, 1992.

\bibitem{Anderson:Catastrophe}
P.~W. {Anderson}.
\newblock {Infrared Catastrophe in Fermi Gases with Local Scattering
  Potentials}.
\newblock {\em Physical Review Letters}, 18:1049--1051, June 1967.

\bibitem{Bettelheim:Kaplan:Wiegmann:Bosons}
E.~{Bettelheim}, Y.~{Kaplan}, and P.~B. {Wiegmann}.
\newblock {Gradient Catastrophe and Fermi Edge Resonances in Fermi Gas}.
\newblock {\em ArXiv e-prints/1011.1993}, November 2010.

\bibitem{wires}
S.~{de Franceschi}, R.~{Hanson}, W.~G. {van der Wiel}, J.~M. {Elzerman}, J.~J.
  {Wijpkema}, T.~{Fujisawa}, S.~{Tarucha}, and L.~P. {Kouwenhoven}.
\newblock {Out-of-Equilibrium Kondo Effect in a Mesoscopic Device}.
\newblock {\em Physical Review Letters}, 89(15):156801--+, September 2002.

\bibitem{Deift:Its:Zhou:Riemann}
P.~{Deift}, A.~{Its}, and X.~{Zhou}.
\newblock {A Riemann-Hilbert approach to asymptotic problems arising in the
  theory of random matrix models, and also in the theory of integrable
  statistical mechanics}.
\newblock {\em Ann. of Math.}, 146:149--235, 1997.

\bibitem{Bettelheim:Abanov:Wiegmann:QHydrodynamics:JPHYSA}
E.~{Bettelheim}, A.~G. {Abanov}, and P.~B. {Wiegmann}.
\newblock {FAST TRACK COMMUNICATION: Quantum hydrodynamics and nonlinear
  differential equations for degenerate Fermi gas}.
\newblock {\em Journal of Physics A Mathematical General}, 41:2003--+, October
  2008.

\bibitem{Combescot:Tanguy:I}
M.~{Combescot} and C.~{Tanguy}.
\newblock {Absorption-edge singularities for a nonequilibrium Fermi sea. I.
  Second-order perturbation theory}.
\newblock {\em Phys. Rev. B}, 50:11484--11498, October 1994.

\bibitem{Combescot:Tanguy:II}
C.~{Tanguy} and M.~{Combescot}.
\newblock {Absorption-edge singularities for a nonequilibrium Fermi sea. II.
  Second-order diagrammatic expansion}.
\newblock {\em Phys. Rev. B}, 50:11499--11507, October 1994.

\bibitem{Combescot:Tanguy:III}
C.~{Tanguy} and M.~{Combescot}.
\newblock {Absorption-edge singularities for a nonequilibrium Fermi sea. III.
  Determinantal nonperturbative theory}.
\newblock {\em Phys. Rev. B}, 52:11698--11710, October 1995.

\bibitem{Levitov:Abanin}
D.~A. {Abanin} and L.~S. {Levitov}.
\newblock {Tunable Fermi-Edge Resonance in an Open Quantum Dot}.
\newblock {\em Physical Review Letters}, 93(12):126802--+, September 2004.

\bibitem{Gutman:Gefen:Mirlin:Toeplitzd}
D.~B. {Gutman}, Y.~{Gefen}, and A.~D. {Mirlin}.
\newblock {Non-equilibrium 1D many-body problems and asymptotic properties of
  Toeplitz determinants}.
\newblock {\em ArXiv e-prints}, October 2010.

\bibitem{Deift:Zhou:Steepest:Descent}
P.~{Deift} and X.~{Zhou}.
\newblock {A steepest descent method for oscillatory Riemann-Hilbert problems}.
\newblock {\em ArXiv Mathematics e-prints}, December 1992.

\bibitem{Sato:Miwa:Jimbo:Impenetrable}
M.~{Jimbo}, T.~{Miwa}, Y.~{M{\^o}ri}, and M.~{Sato}.
\newblock {Density matrix of an impenetrable Bose gas and the fifth
  Painlev{\'e} transcendent}.
\newblock {\em Physica D Nonlinear Phenomena}, 1:80--158, April 1980.

\bibitem{Its:Korepin:DiffEqForCorr}
A.~R. {Its}, A.~G. {Izergin}, V.~E. {Korepin}, and N.~A. {Slavnov}.
\newblock {Differential Equations for Quantum Correlation Functions}.
\newblock {\em International Journal of Modern Physics B}, 4:1003--1037, 1990.

\end{thebibliography}

\end{document}